# Spectral analyses of *trans*- and *cis*-DOCO transients via comb spectroscopy

Thinh Q. Bui[1], P. Bryan Changala[1], Bryce J. Bjork[2], Qi Yu[3], Yimin Wang[4], John F. Stanton[5], Joel Bowman[3], Jun Ye[1]

[1] *JILA, National Institute of Standards and Technology, and Department of Physics, University of Colorado, Boulder, CO 80309, USA*

[2] *Honeywell International, 303 Technology Ct., Broomfield, Colorado, 80021, USA*

[3] *Cherry L. Emerson Center for Scientific Computation and Department of Chemistry, Emory University, Atlanta, Georgia 30322, United States*

[4] *2604 Kings Lake Court NE, Atlanta, Georgia 30345, USA*

[5] *Department of Chemistry, University of Florida, Gainesville, Florida 32611, USA*

*Abstract*

We use time-resolved direct frequency comb spectroscopy in the mid-infrared to obtain high-resolution rovibrational spectra of products produced from the OD+CO reaction. In this work, we present spectral analyses for isotopologues of the transient DOCO radicals from this reaction in the OD stretch region. The analyses were performed with aid of two different theoretical approaches based on both perturbation theory and variational calculations used for prediction of rovibrational spectra of polyatomic molecules. We discuss the advantages and challenges of our current approach for studying spectroscopy and dynamics of transient molecules.

# 1. Introduction

The use of time-resolved spectroscopy for the study of elementary reaction processes, a key driver in the fundamental understanding of chemical reaction mechanisms and molecular dynamics [1], has experienced revolutionary transformation beginning from Norrish and Porter's seminal flash photolysis experiment to ultrafast "femtochemistry" by Ahmed Zewail [2]. The development of ultrafast lasers served as a cornerstone for this transition. Taking a different path, high resolution spectroscopy and precision measurement have motivated the development of stable lasers and frequency-domain approaches. The great merge of these two scientific paths led to the eventual development of the optical frequency comb [3]. The frequency comb possesses broad spectral bandwidth and high spectral resolution in the frequency domain, making it a suitable light source for high-resolution spectroscopy in what has been termed "direct frequency comb spectroscopy" (DFCS) [4]. The versatility of DFCS has more recently been extended to studies of high resolution spectroscopy of large molecules [5,6] and chemical kinetics [7-10]. Continuing efforts are focused towards construction of high power frequency comb sources that cover 5 to 10 μm for future advances in high resolution molecular spectroscopy and dynamics [11].

In Ref. [7] we reported the use of cavity-enhanced direct frequency comb spectroscopy to determine the real-time kinetics of the OD+CO reaction, which is important in atmospheric and combustion chemistry [12]. When combined with a dispersive spectrometer, this technique achieves the time resolution necessary for monitoring real-time formation and decay of the reaction intermediate (DOCO) and product ($CO_2$) from the OD+CO reaction. Here, we provide a detailed presentation on the spectral analyses of the reaction products, specifically the intermediate transients of *trans*- and *cis*-DOCO, based on high resolution spectroscopy data in the OD stretch band region (λ ~ 3.7 to 4.2 μm). These spectroscopic investigations form an important part of our understanding of the structure and dynamics of these transient species [13].

High-resolution spectra of H(D)OCO intermediates motivate the development of a more accurate OH(D)+CO global potential energy surface (PES), especially in the low energy regions probed by observation of vibrational fundamentals in the infrared wavelengths [14,15,16]. Relying on ab initio PES, recent theoretical work has focused on the dissociation dynamics of H(D)OCO isomers to OH(D)+CO or H(D)+$CO_2$ products [15,17]. Gas phase infrared spectroscopy of *trans*-H(D)OCO have now been reported for the $\nu_1$ O-H(D) stretch [18-20] and $\nu_2$ C=O stretch [21] vibrational fundamental modes. Extensive microwave and millimeter-wave studies for the vibrational ground state have also been reported for both *trans* and *cis* isomers [22-24]. The spectroscopic data reported in this work thus provides important information for a more comprehensive understanding of the structure of DOCO isomers.

## 2. Methods

### I. Transient DOCO production

Detailed descriptions of the DOCO-forming reaction processes have been described in our previous work [9]; only a brief review will be given here. First, $O_3$ gas is photo-dissociated in a room temperature reaction cell (continuous gas flow) by a 266 nm pulse (10 ns, 35 mJ/pulse) to produce $O(^1D)$ and $O_2$. In the presence of $D_2$, the reaction of $O(^1D)+D_2$ produces OD radicals. CO is then added to initiate the OD+CO reaction, which produces reactive intermediates *cis*- and *trans*-DOCO and product $CO_2$. This work will focus on the high-resolution spectroscopy and supporting rovibrational calculations for the DOCO intermediates (*trans*-DO$^{12}$CO, *trans*-DO$^{13}$CO, *cis*-DO$^{12}$CO, and *cis*-DO$^{13}$CO) in the OD(v=1) stretch region.

## II. Time-resolved frequency comb spectroscopy

The mid-IR frequency comb light is produced from a tunable ($\lambda \sim$ 3 to 5 μm) optical parametric oscillator (OPO) synchronously pumped with an 10 W ytterbium fiber comb ($\lambda \sim$ 1.06 μm) [25]. In this work, the OPO wavelength is tuned from 3.6 μm to 4.3 μm (2300 to 2800 cm$^{-1}$, average power ~ 200 to 500 mW). The repetition rate ($f_{rep}$) of the comb is ~ 136.7 MHz.

Light from the OPO is sent into an optical cavity (which also served as the reaction cell) enclosed by two high reflectivity mirrors for cavity-enhanced absorption spectroscopy. The measured finesse of the cavity is shown in Fig. 1A. The length of the cavity is approximately 54.9 cm resulting in a cavity free spectral range (FSR) of ~ 273 MHz, or 2×$f_{rep}$. Therefore, every other comb mode is coupled into the cavity. In this experiment, photo-dissociation of O$_3$ by the Nd:YAG laser causes a transient increase in the gas pressure, which changes the effective optical path length. In the case of a tight comb-cavity locking method like the PDH lock, this sudden disturbance results in lock instability and/or introduces cavity transmission noise (frequency-to-amplitude noise conversion). Therefore, to maintain cavity transmission, the swept cavity lock method [26] was used rather than the PDH lock method. Here, the comb $f_{rep}$ is modulated at 50 kHz, and the cavity transmission signal is demodulated and fed back to the cavity piezo to keep the cavity FSR locked to the $f_{rep}$ of the comb laser. At the expense of lowering the laser-cavity coupling duty cycle (decreased cavity transmission), the swept cavity lock technique has the advantage of being less sensitive to the photolysis process.

The transmitted comb spectrum is dispersed by a spectrometer that comprises a combination of a virtually-imaged phased array (VIPA) etalon [27] and a reflective diffraction grating. Since the cavity-filtered comb mode spacing (273 MHz) cannot be resolved by the VIPA etalon, the spectrometer sets the resolution (~ 900 MHz) rather than the linewidth of the comb mode (~ 50 kHz). Output from the VIPA etalon (vertical dispersion) is cross-dispersed with a grating (horizontal dispersion), and imaged on an InSb camera. The integration time of the camera sets the time resolution ($t_{int} \geq 10$ μs). This configuration allows for simultaneous measurement of approximately 65 cm$^{-1}$ of the comb spectral width at a VIPA limited resolution of ~ 900 MHz (~ 0.03 cm$^{-1}$). This corresponds to more than 2000 spectrally resolved elements that are acquired simultaneously within 10 μs. The experiments are conducted at ~100 Torr and room temperature, which means that the combined Doppler and pressure broadened lineshape exceeds the VIPA limited resolution. Thus, the experimental conditions for studying the OD+CO reaction ultimately determine the spectral resolution, not the frequency comb or spectrometer.

The dispersive spectrometer provides the necessary time resolution to observe the short-lived (100 μs) DOCO intermediates via a pump–comb probe experiment with the Nd:YAG photolysis (pump) laser. The integration of the cavity transmitted comb light on the InSb camera is synchronized with the photolysis pulse. To obtain a direct absorption signal, rapid successive acquisitions of the reference 'R' (pre-photolysis) and signal 'S' (post-photolysis) camera images are recorded. The experimentally chosen temporal separation between the R and S defines the reaction kinetics time. The absorbance is determined from

$$A = -ln\left(\frac{S-B}{R-B}\right). \quad\quad\quad \text{eq. 1}$$

In eq. 1, 'B' refers to the background camera image with the IR beam blocked by a mechanical shutter, which is measured 4 ms before the R image. Fast subtraction of the B image from both the R and S mitigates additional noise caused by the temperature-dependent dark current offset drifts of the InSb camera.

The duty cycle of the experiment is limited by the 10 Hz repetition rate of the Nd:YAG laser. The 100 ms separation between photolysis pulses provides more than sufficient time for gas pump out (residence time is ~ 20 ms). Due to latency in the acquisition software, the actual acquisition repetition rate is approximately 3 Hz. The single shot absorption sensitivity is estimated by the noise-equivalent absorption (NEA) per spectral element (to normalize the comb bandwidth), which is given by

$$\text{NEA} = \sigma_A \frac{\pi}{FL_p}\sqrt{\frac{T}{M}}. \qquad \text{eq. 2}$$

Here, $\sigma_A$ is the standard deviation in the single shot absorbance calculated by eq. 1, $F$ is the cavity finesse, $L_p$ is the photolysis pathlength, $T$ is the total period for the measurement of $A$, and $M$ is the number of resolvable spectral elements per camera image. At peak finesse, NEA is $2\times10^{-10}$ cm$^{-1}$ Hz$^{-1/2}$, which is a factor of five better than the previous time-resolved frequency comb experiment with reported a NEA of $1.1\times10^{-9}$ cm$^{-1}$ Hz$^{-1/2}$ [10]. The improvement can be attributed to higher cavity finesse in the current apparatus.

To further enhance the absorption sensitivity, many single shot spectra are averaged. Fig. 1B shows the Allan deviation of the absorbance at two different spectral baseline points as a function of averaging time $\tau$. Here, the baseline noise averages down as $\tau^{-1/2}$ even after 30 minutes of averaging (at a 3 Hz acquisition rate). This observation reveals the additional noise reduction advantage of our 10× faster differential measurement compared to a previous study [5].

### III. Rovibrational calculations

#### A. MULTIMODE

We have performed vibrational self-consistent field/virtual state configuration interaction (VSCF/VCI) calculations, as implemented in the code MULTIMODE (MM) [28]. The exact Watson Hamiltonian is used in the representation of mass-scaled normal modes. For all the cases, we use the 6-mode representation of the potential (no approximation made). The two different potential energy surfaces used are developed by Chen *et al.* [29] and Wang *et al.* [16] The former PES is a global surface starting from the OH+CO asymptote to the H+$CO_2$ product, while the latter is centered around the minima of the *trans-* and *cis-*HOCO isomers. For each normal mode, twenty-two Gauss quadrature points are selected in generating a set of harmonic basis functions. In VCI calculations, the sum of mode excitations of all 6 normal modes are 14, 14, 13, 13, 11,10 for 1-mode to 6-mode excitations. The final size of VCI matrix is 20877.

B. VPT2

Separately, we also have performed second-order vibrational perturbation theory (VPT2) [30] calculations for both the $^{12}$C and $^{13}$C isotopologues of the *cis* and *trans* isomers. Standard semi-diagonal quartic force fields with respect to the rectilinear normal coordinates are calculated at the frozen-core CCSD(T) level of theory with the ANO1 basis set [31] using the CFOUR package [32]. The VPT2 predictions include both anharmonic vibrational frequencies and vibrational corrections to the rotational constants.

## 3. Results and discussions

The calculated DOCO vibrational frequencies obtained using MULTIMODE and VPT2 are compiled in Tables 1 and 2, respectively. For *trans*-DOCO, the $^{13}$C substitution is not anticipated to significantly shift the origin of the OD stretch band relative to $^{12}$C, as corroborated by the nearly identical computed values for the two carbon isotopologues. For *cis*-DOCO, the predicted values for both the $^{12}$C and $^{13}$C isotopologues provide guidance for our search for the *cis*-DOCO radical in the OD stretch vibrational band. For this purpose, MULTIMODE using two different PES and VPT2 all provided good agreement for the OD stretch frequency within ~ 10 cm$^{-1}$.

The absorption spectrum of each major species produced from the OD+CO reaction is shown in Fig. 2. The large bandwidth of the high reflectivity mirrors spans a measurement range of 2380 cm$^{-1}$ to 2760 cm$^{-1}$ (3.6 μm to 4.2 μm), which allows us to measure $CO_2$, $DO_2$, $D_2O$, OD, *cis*- and *trans*-DOCO. The simulated OD and $D_2O$ line positions are obtained from Abrams *et al.* [33] and Toth *et al*. [34], respectively. The $DO_2$ spectrum is simulated from measured rovibrational constants from Lubic *et al*. [35] All spectra are simulated at T = 295 K, including that of the DOCO isomers, which, despite being produced with significant chemical activation from the OD+CO reaction, are rapidly thermalized to room temperature by high background concentrations of $N_2$ and CO.

I.   *trans*-DOCO

*trans*-DOCO is a planar, near-prolate asymmetric top. The ratio of the a-type to b-type integrated band intensities for its OD stretch fundametnal is estimated to be $|\mu_a/\mu_b|^2$ ~ 3 based on jet-cooled spectra of *trans*-HOCO [20]. Despite the similar band intensities, previous room temperature vibrational spectra in the OH(D) stretch region of *trans*-H(D)O$^{12}$CO are dominated by a-type transitions with no apparent signatures of b-type transitions [18,19], which is consistent with our own measurements. For the *trans*-DO$^{12}$CO isotopologue, both of ground state [23,24] and excited OD(v=1) stretch [18] rovibrational constants have been previously reported, so we will not discuss that here.

We report infrared spectroscopy of the *trans*-DO$^{13}$CO isotopologue, for which no previous reports have been made. Figure 3 show the experimental and fitted (inverted) spectra for *trans*-DO$^{13}$CO. The fits utilize parameters for the Watson A-reduced effective

Hamiltonian (I$^r$ representation) and are performed using PGOPHER [36]. Since there is no previous measurement of this vibrational band, the previously described VPT2 calculations provided initial guesses for both the vibrational band origin and rotational constants (A, B, C).

The rotational energies of a near-prolate asymmetric top increase approximately as (A-(B+C)/2)$K_a^2$ [the $K_a$ quantum number is the projection of the rotational angular momentum along the principal a-axis]. The propensity rule for a-type transitions is $\Delta K_a$=0, while for b-type transitions it is $\Delta K_a = \pm 1$. The rotational constant A, which largely determines the spacing between different $K_a$ stacks, is poorly constrained in an a-type spectrum because of the $\Delta K_a$=0 propensity rule. By only observing a-type transitions in this experiment, the strong correlation in the fitted values of A, B, and C precludes their accurate individual determination. Thus the values for the rotational constant A for the ground ($A_0$) and vibrationally excited state ($A_v$) are determined by correcting the experimental $^{12}$C values with the calculated VPT2 isotopic shift and fixed in the fit. The quartic centrifugal distortion terms are fixed to the experimental *trans*-DO$^{12}$CO values reported by Petty and Moore [18] for both carbon isotopologues. The instrument and pressure-broadened transitions (linewidth ~ 0.03 cm$^{-1}$) for the DOCO isomers cannot resolve the asymmetry doubling (for levels $K_a \neq 0$), which constrains the difference in B and C. Therefore, only the average value of (B+C)/2 for the ground ($B_0$ and $C_0$) and vibrationally excited states ($B_v$ and $C_v$) are fitted for *trans*-DO$^{13}$CO, along with the v=1 band origin.

The fitted *trans*-DO$^{13}$CO rovibrational constants are compiled in Table 3. The standard deviation of the fit is ~ 0.013 cm$^{-1}$, which is well below the uncertainty of ~ 0.1 cm$^{-1}$ for the experimental transition energies. The observed agreement between the measured and predicted spectra demonstrates that only a few free parameters (the average of the B and C rotational constants and the band origin) are required to reproduce the pressure-broadened, room temperature spectrum of *trans*-DO$^{13}$CO to within experimental uncertainty.

## II. *cis*-DOCO

Prior to our work, there have been no previous reports of the gas phase vibrational spectrum of the *cis*-H(D)OCO isomer. Pure rotational microwave spectra have been reported by Oyama *et al*. [37] and McCarthy *et al*. [22], both of whom used an electric discharge source to produce the *cis* isomer. These measurements provide detailed structural information and form an excellent starting point for the present infrared experiments. In particular, by adding calculated VPT2 vibrational and isotopic shifts to the measured ground state rotational constants we obtain a reasonable predicted spectrum of the OD stretch band for fitting to our experimental spectrum.

In the *cis*-DOCO molecule (also a planar, near-prolate asymmetric top), the strongest OD stretch transition dipole component is aligned along the b-axis. VPT2 calculations predict that the ratio of a-type to b-type integrated band intensities for *cis*-DOCO is ~ 0.077. The observed spectrum (Figure 4) is dominated by b-type transitions (propensity rule of $\Delta K_a = \pm 1$), signified by prominent, yet unresolved, Q branch transitions each originating from the ground state rotational levels of a given $K_a$ value. Here, only $K_a$ = 1-8 transitions are unambiguously identified. For a near-prolate asymmetric top, the frequency spacing between the Q branches of neighboring $K_a$ sub-bands is approximately given by A-(B+C)/2, or ~ 6.6 cm$^{-1}$. a-type transitions were not observed here for either carbon isotopologues of *cis*-DOCO.

Since b-type transitions require a change in the $K_a$ quantum number, the rotational constant A can be obtained by fitting the $K_a$ stack energy spacing. Because the ground state rotational constants for the $^{12}$C isotopologue have been measured by microwave spectroscopy [22,23], only the excited state value of $A_v$ and the band origin are fitted. For the $^{12}$C isotopologue, the initial guess for $A_v$ is obtained from the calculated vibrational shift to the measured ground state rotational constant $A_0$ provided by VPT2. For the $^{13}$C isotopologue, the initial guess for $A_v$ is determined from the sum of the computed vibrational and isotopic shifts to the measured $A_0$ for the $^{12}$C isotopologue. For both the ground and excited state, the quartic centrifugal distortion terms are fixed to the ground state values measured by McCarthy *et al*. [22]. As shown in Figure 4, fitting with only the $A_v$ constant and band origin is sufficient to match the experimental spectrum to within experimental uncertainty. Here, the standard deviation of the fit is 0.014 cm$^{-1}$. The fitted rovibrational constants for *cis*-DOCO are compiled in Table 3.

Comparison of the measured *cis*-DOCO band origin reveal nearly exact agreement (~ 1 cm$^{-1}$) with the values computed by MULTIMODE using the global PES by Chen *et al.* [29] (Table 1). The predicted band origins obtained by Guo and co-workers [38] and our MULTIMODE calculation using the PES by Wang *et al.* [16] (Table 2) achieved similar agreement to within ~ 5 cm$^{-1}$ of the measured value. The largest discrepancy in the predicted band origin of ~ 15 cm$^{-1}$ is observed using VPT2 (Table 3), even though the predicted vibrational and isotopic shifts to the rotational constants from VPT2 are accurate to within ~ 100 MHz (well below the experimental resolution) of the measured values. Finally, we note that these state-of-the-art theoretical methods accurately capture the subtle anharmonic effects that give rise to the small isotopic shifts (~ 0.2 cm$^{-1}$) in the vibrational transition frequencies for both DOCO isomers.

## 4. Conclusion

In this work, we report the high resolution (~ 0.03 cm$^{-1}$) spectroscopy of the isotopologues of DOCO isomers from the OD+CO reaction in the mid-IR (3.7 to 4.2 μm). Using time-resolved frequency comb spectroscopy, we have reported spectra and partial rovibrational analyses for the OD stretch bands of *cis*-DO$^{12}$CO, *cis*-DO$^{13}$CO, and *trans*-DO$^{13}$CO. A future direction of research is the development of high power frequency comb sources in both the 5 to 7 μm and 8 to 10 μm wavelength regions, which cover the carbonyl (C=O) stretch and D-O-C bend mode vibrational frequencies, respectively, of both DOCO isomers. These two vibrational bands have significantly larger intensities than the OD stretch mode. Sears *et al.* [21] reported the only measurement of the carbonyl stretch of gas phase *trans*-DOCO, but this mode has not been seen for the *cis* isomer. These measurements at longer infrared wavelengths will provide useful spectroscopic parameters to improve the quality of the global PES used to accurately model OH(D)+CO reaction kinetics and dynamics. Finally, we hope that our work will motivate further studies at even higher resolution (microwave or infrared) to complete our partial rotational analyses of these H(D)OCO radicals, ideally at much lower temperatures and pressures.

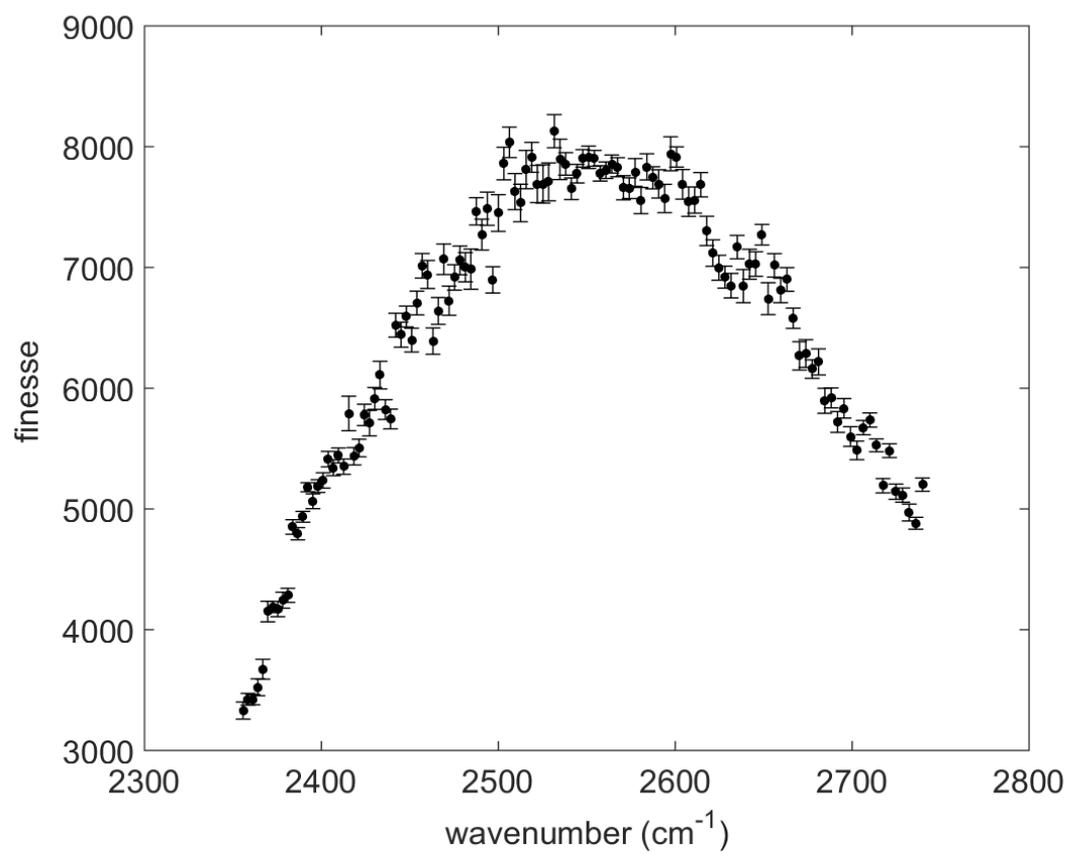

**Figure 1A: Cavity finesse.** The cavity finesse for a cavity length of 54.9 cm was obtained using the cavity ringdown technique.

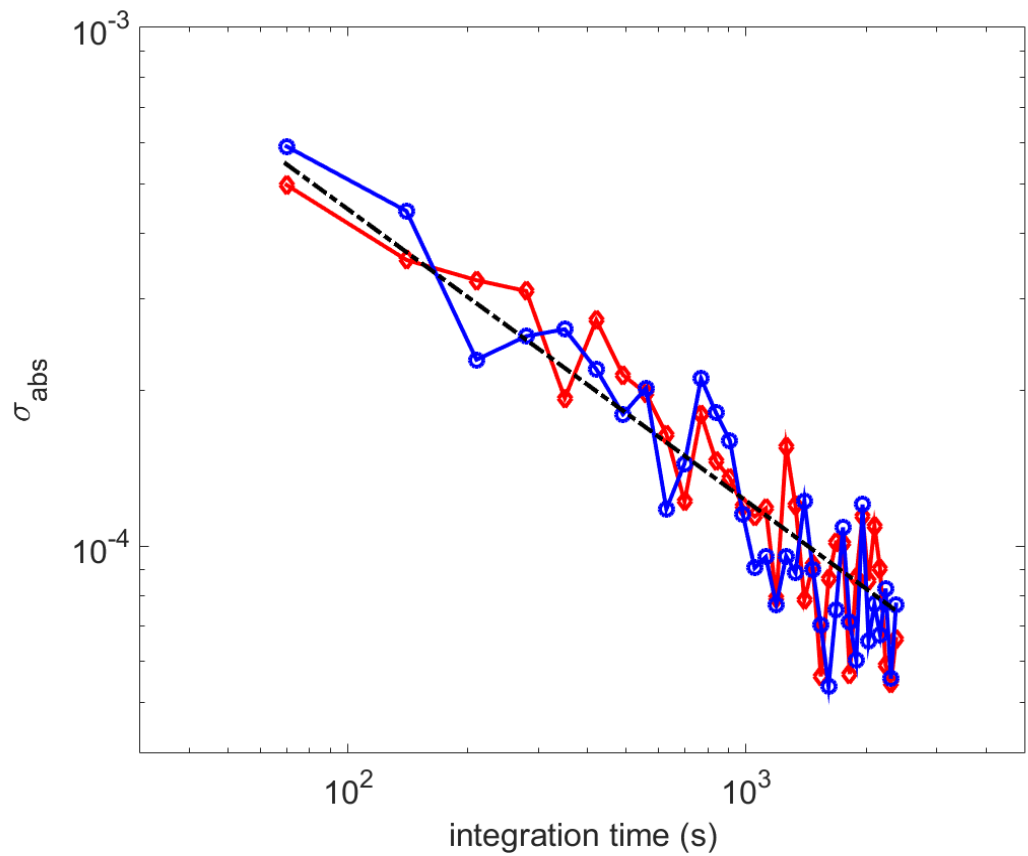

**Figure 1B:** Allan deviation of the absorbance determined from eq. 1. Blue and red traces correspond to the measured absorbance at two different baseline points in the spectrum. The black dashed line is the $\tau^{1/2}$ dependence, where $\tau$ is the averaging time.

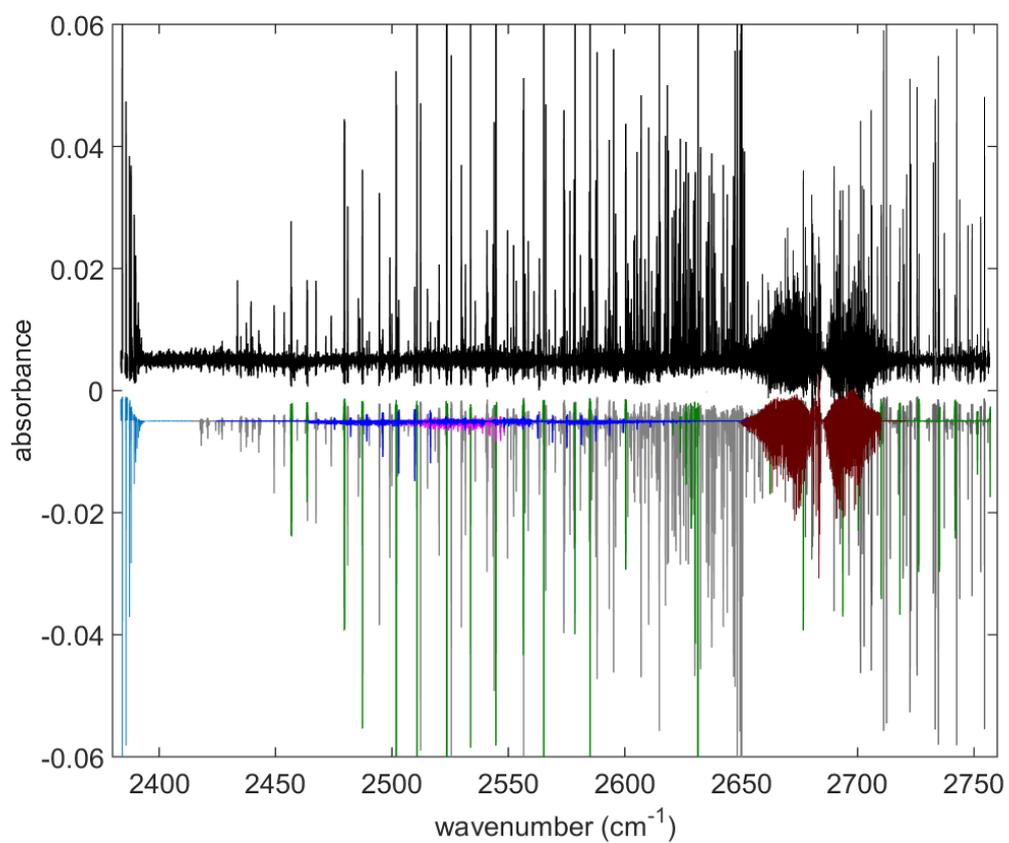

**Figure 2: Spectral survey of all species from the OD+CO reaction. Cyan: $CO_2$; blue: *cis*-$DO^{12}CO$; brown: *trans*-$DO^{12}CO$; pink: $DO_2$; green: OD; gray: $D_2O$.**

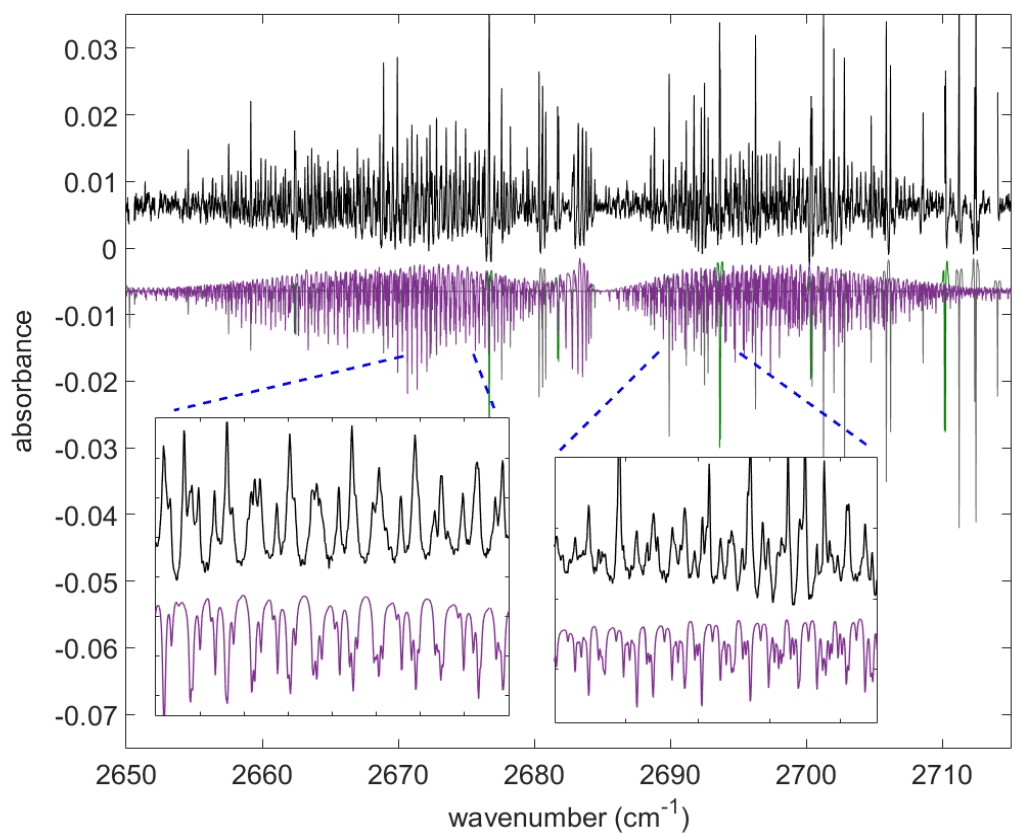

**Figure 3:** Experimental *trans*-DO$^{13}$CO spectrum and fit (inverted). In the insets, the simulated OD (green) and D$_2$O (gray) lines have been removed for clarity.

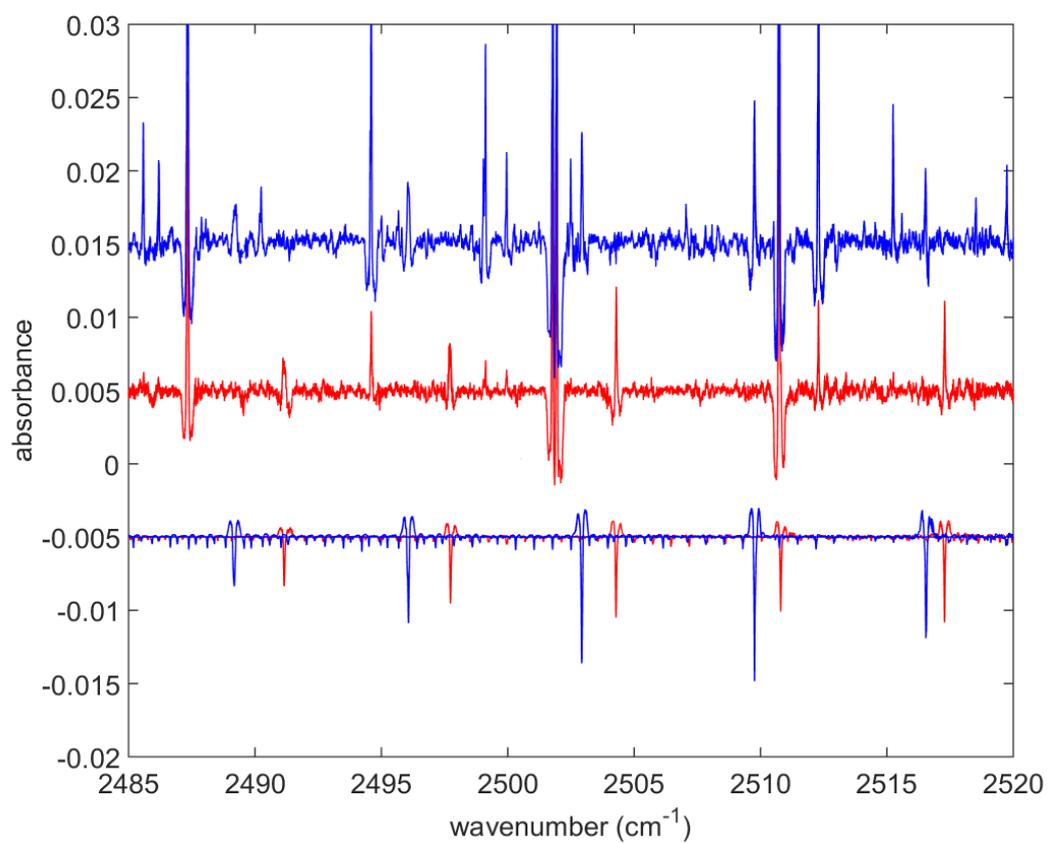

**Figure 4: Carbon isotopologues of *cis*-DOCO. blue: *cis*-DO$^{12}$CO; red: *cis*-DO$^{13}$CO. The simulated spectra for *cis*-DOCO are shown inverted. In the inverted spectra, the simulated OD and D$_2$O lines have been excluded for clarity.**

**Table 1: Vibrational frequencies using MULTIMODE**

| Mode | cis-DO$^{12}$CO[c] | cis-DO$^{13}$CO[c] | trans-DO$^{12}$CO[c] | trans-DO$^{13}$CO[c] |
|---|---|---|---|---|
| 1 torsion | 451.57 | 447.26 | 396.72 | 393.92 |
| 2 O-C-O bend | 535.12 | 530.95 | 589.11 | 582.64 |
| 3 H-O-C bend | 957.22 | 955.59 | 902.55 | 900.86 |
| 4 C-O stretch | 1116.05 | 1091.19 | 1083.79 | 1063.13 |
| 5 C=O stretch | 1818.27 | 1777.48 | 1851.55 | 1813.78 |
| 6 O-D stretch | 2540.93 | 2540.77 | 2686.2 | 2686.25 |

[c]PES from Chen *et al.* [29]

| Mode | cis-DO$^{12}$CO[d] | cis-DO$^{13}$CO[d] | trans-DO$^{12}$CO[d] | trans-DO$^{13}$CO[d] |
|---|---|---|---|---|
| 1 torsion | 460.51 | 455.95 | 392.35 | 389.59 |
| 2 O-C-O bend | 536.97 | 532.73 | 587.86 | 581.36 |
| 3 H-O-C bend | 953.91 | 952.69 | 900.55 | 898.95 |
| 4 C-O stretch | 1118.85 | 1093.81 | 1083.95 | 1063.15 |
| 5 C=O stretch | 1820.93 | 1780.43 | 1852.83 | 1814.09 |
| 6 O-D stretch | 2544.57 | 2544.4 | 2685.54 | 2685.91 |

[d]PES from Wang *et al.* [16]

**Table 2: Vibrational frequencies using CCSD(T)/ANO1 VPT2 (in cm$^{-1}$)**

| Mode | cis-DO$^{12}$CO | cis-DO$^{13}$CO | trans-DO$^{12}$CO | trans-DO$^{13}$CO |
|---|---|---|---|---|
| 1 torsion | 453.6757 | 449.1907 | 394.5547 | 392.2696 |
| 2 O-C-O bend | 535.0366 | 530.9112 | 587.5528 | 581.4452 |
| 3 H-O-C bend | 963.4104 | 958.6848 | 899.2502 | 898.1702 |
| 4 C-O stretch | 1121.2705 | 1095.9654 | 1080.4844 | 1059.739 |
| 5 C=O stretch | 1814.7033 | 1773.6179 | 1847.3807 | 1783.3468 |
| 6 O-D stretch | 2555.4755 | 2555.2693 | 2688.1609 | 2688.3992 |

**Table 3:** *trans*-DOCO and *cis*-DOCO molecular constants (in MHz)

| Parameter | *trans*-DO$^{12}$CO | *trans*-DO$^{13}$CO | *cis*-DO$^{12}$CO | *cis*-DO$^{13}$CO |
|---|---|---|---|---|
| $\nu_0$ (cm$^{-1}$) | 2684.1[a] | 2684.159(2) | 2539.909(3) | 2539.725(4) |
| $A_0$ | 154685.537[b] | 148175.6034 | 110105.52[b] | 106124(5) |
| $B_0$ | 10685.952[b] | --- | 11423.441[b] | 11420.075 |
| $C_0$ | 9981.624[b] | --- | 10331.423[b] | 10291.999 |
| $(B_0+C_0)/2$ | 10333.788[b] | 10310(1) | 10877.432[b] | 10856.037 |
| $A_v$ | 153431.4[a] | 147030.0607 | 109313(4) | 105423(5) |
| $B_v$ | 10671.06[a] | --- | 11422.882 | 11419.559 |
| $C_v$ | 9963.386[a] | --- | 10324.228 | 10284.951 |
| $(B_v+C_v)/2$ | 10317.223[a] | 10293(1) | 10873.555 | 10852.255 |

a Petty and Moore [18]
b. McCarthy *et al.* [22]

## 6. Acknowledgements

We thank H. Guo for stimulating discussions. We acknowledge financial support from AFOSR, DARPA SCOUT, NIST, and NSF JILA Physics Frontier Center. J.F. Stanton acknowledges financial support from the U.S. Department of Energy, Office of Basic Energy Sciences for Award DE-FG02-07ER15884. J. M. Bowman thanks the National Science Foundation (CHE-1463552) for financial support. T. Q. Bui is supported by the National Research Council Research Associate Fellowship, P. B. Changala is supported by the NSF GRFP.